\documentclass[twocolumn,prc,superscriptaddress,unsortedaddress,showpacs,preprintnumbers,amsmath,amssymb]{revtex4}

\usepackage{graphicx}
\usepackage{dcolumn}
\usepackage{bm}
\usepackage{txfonts}

\def\beq{\begin{equation}}
\def\eeq{\end{equation}}
\def\bea{\begin{eqnarray}}
\def\eea{\end{eqnarray}}

\def\fun#1#2{\lower3.6pt\vbox{\baselineskip0pt\lineskip.9pt
  \ialign{$\mathsurround=0pt#1\hfil##\hfil$\crcr#2\crcr\sim\crcr}}}

\begin{document}
\preprint{}

\title{Density-dependent symmetry energy at subsaturation densities from nuclear mass differences}

\author{Xiaohua Fan}\affiliation{Institute of Modern Physics, Chinese
Academy of Sciences, Lanzhou 730000, China} \affiliation{University
of Chinese Academy of Sciences, Beijing, 100049, China}
\author{Jianmin Dong}\email[ ]{djm4008@126.com}\affiliation{Institute of Modern Physics, Chinese
Academy of Sciences, Lanzhou 730000, China}\affiliation{State Key
Laboratory of Theoretical Physics, Institute of Theoretical Physics,
Chinese Academy of Sciences, Beijing, 100190, China}
\author{Wei Zuo}
\affiliation{Institute of Modern Physics, Chinese Academy of
Sciences, Lanzhou 730000, China}\affiliation{State Key Laboratory of
Theoretical Physics, Institute of Theoretical Physics, Chinese
Academy of Sciences, Beijing, 100190, China}
\date{\today}

\begin{abstract}
We extract the mass-dependent symmetry energy coefficients
$a_{\text{sym}}({A})$ with the nuclear mass differences reducing the
uncertainties as far as possible. The estimated
$a_{\text{sym}}({A})$ of $^{208}\text{Pb}$ is $22.4\pm 0.3 $ MeV,
which is further used to analyze the density-dependent nuclear
matter symmetry energy at subsaturation densities. The slope
parameter of the symmetry energy at the saturation density
$\rho_{0}$ is $L=50.0\pm15.5$ MeV. Furthermore, it is found that, at
the density of $\rho=0.69\rho_{0}=0.11$fm$^{-3}$, the symmetry
energy $S(\rho=0.11\text{fm}^{-3})=25.98\pm0.01$ MeV and the
correspondingly slope parameter is $L=49.6\pm6.2$ MeV, which are
consistent with other independent analysis.
\end{abstract}


\maketitle

The equation of state (EOS) of isospin asymmetric nuclear matter is
an active research field at present because of its importance in
nuclear physics and in particular in astrophysics. Unfortunately,
the variation of the EOS with respect to baryon density is still
being intensely debated, especially the symmetry energy which
characterizes its isospin-dependence. The density-dependent symmetry
energy plays a crucial role in understanding a variety of issues in
nuclear physics and astrophysics, such as the heavy ion reactions
\cite{VB,AWS,BAL,JML}, the stability of superheavy nuclei \cite{JD},
the structures, composition and cooling of neutron stars
\cite{NS1,BKSP,NS,BG}. Because of its great importance, many authors
concentrate on this issue within many independent approaches, such
as the heavy ion collision, microscopic and phenomenological nuclear
many body theories and collective excitations. At current one grasps
some basic knowledge about the symmetry energy at low densities,
while at high densities one almost know nothing even its variation
tendency as the density. The slope parameter $L$ governing the
density dependence of $S(\rho )$ around the saturation density
$\rho_{0}$, has been found to correlate linearly with the neutron
skin thickness of heavy nuclei such as $^{208}$Pb \cite{BAB,ST,RJF}.
Therefore, a measurement of $\Delta R_{np}$ with a high accuracy is
a strong constraint of the density dependence of symmetry energy at
subnormal densities. Due to the large uncertainties in measured
neutron skin thickness, this has not been possible now. However, we
may constrain the symmetry energy effectively with the help of other
approaches.

Recently, many independent investigations have been performed to
constrain the density dependence of the symmetry energy. A detailed
summary of the recent progress can be found in Ref. \cite{MBT2} and
the introduction of Ref. \cite{dong13}. Lately, Agrawal {\it et al.}
calculated the density distributions in both spherical and well
deformed nuclei within microscopic framework with different energy
density functionals giving $L = 59.0 \pm 13.0$ MeV \cite{AAA}. Dong
{\it et al.} probed the density dependence of the symmetry energy
around the saturation density with the $\beta^{-}$-decay energies of
odd-A heavy nuclei \cite{dong13}, and obtained $L=50\pm15$ MeV. Wang
and Li observed that a linear relationship between the $L$ and the
root-mean square (rms) charge radius difference of the
$^{30}$S-$^{30}$Si mirror pair, and the estimated slope parameter is
about $L = 54 \pm 19$ MeV from the coefficient of their proposed
charge radius formula \cite{charge}. In this Brief Report, we employ
the nuclear mass differences to derive the symmetry energy
coefficient $a_{sym}(A)$ of heavy nuclei and then explore the
density dependence of the nuclear matter symmetry energy at
subsaturation densities.

We extract the $a_{\text{sym}}({A})$ with the differences of
experimental nuclear mass \cite{Audi} in order to reduce the
uncertainties as far as possible. The binding energy $B(Z,A)$ of a
nucleus can be described by the well-known liquid drop formula
\begin{equation}
B(Z,A)=a_{v}A-a_{s}A^{2/3}-E_{c}-a_{sym}(A)\beta
^{2}A+E_{p}+...\label{AA}
\end{equation}%
The Coulomb energy that includes charge exchange correction is given
by
\begin{equation}
E_{c}=a_{c}\frac{Z(Z-1)}{A^{1/3}(1+\Delta )}\left(
1-0.76Z^{-2/3}\right),
\end{equation}
where the parameter $\Delta$ was introduced to describe the effect
of the Coulomb interaction on the surface asymmetry and the effect
of the surface diffuseness on the Coulomb energy \cite{PD2}, taking
the form
\begin{equation}
\Delta =\frac{5\pi ^{2}}{6}\frac{d^{2}}{r_{0}^{2}A^{2/3}}-\frac{1}{%
1+A^{1/3}/\kappa }\frac{N-Z}{6Z}.
\end{equation}
$d \approx 0.55$ fm \cite{PD2} is the diffuseness parameter in the
Fermi function from the parametrization of nuclear charge
distributions and $r_{0}$ is the nuclear-radius constant satisfying
$3/(4\pi r_{0}^{3})=0.16$ fm$^{-3}$. The meaning of the $\kappa$ is
discussed later. Here the independent variables are mass number $A$
and isospin asymmetry $\beta$. Thus, the proton number is
$Z=A(1-\beta)/2$. The parameter $a_{c}=0.71$ is known very well
\cite{W2}, in particular well determined from the masses of mirror
nuclei \cite{MIR1,MIR2}. Performing a partial derivative of $B(Z,A)$
with respect to the isospin asymmetry $\beta$ in Eq. (\ref{AA}), the
symmetry energy coefficient can be expressed as
\begin{equation}
a_{sym}(A)=-\left( \frac{\partial B(Z,A)}{\partial \beta
}+\frac{\partial E_{c}(Z,A)}{\partial \beta }\right) /\left( 2\beta
A\right).
\end{equation}
Here the partial derivative $\partial B(Z,A)/\partial \beta $ is
replaced by the difference
\begin{equation}
\frac{\partial B(Z,A)}{\partial \beta }\approx \frac{B(Z+1,A)-B(Z-1,A)}{%
\beta _{2}-\beta _{1}},
\end{equation}
where $\beta_{1}$ and $\beta_{2}$ are the isospin asymmetry of
nuclei $(Z-1,A)$ and  $(Z+1,A)$, respectively. The difference of the
$\Delta$ between the two neighboring nuclei is neglected since it is
quite small. The nuclei involving magic numbers are excluded to
avoid the strong shell effects. Because of the two neighboring
nuclei $(Z-1,A)$ and $(Z+1,A)$ sharing the same odevity, the pairing
energy is canceled out in binding energy difference $\Delta
B=B(Z+1,A)-B(Z-1,A)$. This is one of the advantages of the present
approach. The shell correction in the binding energy for the two
neighboring nuclei should be close with each other because the
densities of the energy levels are not expected to change distinctly
considering they share the same odevity. Accordingly, the shell
correction energies to their masses could be canceled to a large
extent leading to a negligible correction to the $\Delta B$. The
contribution of the Coulomb energy is relatively clear, which is the
primary advantage of this approach. This method should be better
than that using the $\beta^{-}$-decay energies $Q_{\beta^{-}}$ of
odd-A heavy nuclei since the nuclear odevity changes in
$\beta$-decay. On the other hand, much more experimental data are
available in the present study compared to that using
$Q_{\beta^{-}}$. The 168 experimental masses of $((Z-1,A), (Z+1,A))$
pairs of translead nuclei are used in the following analysis.

\begin{table}[h]
\label{table2} \caption{Comparison between the $L$ values obtained
in the present work and those from other recently independently
analyses.}
\begin{ruledtabular}
\begin{tabular}{llllllllllllllll}
Reference  & Method  & $L$ (MeV)  \\
\hline
Ref. \cite{ML}  &nuclear masses &  $53 \lesssim L \lesssim 79$   \\

Ref. \cite{QPO} & quasiperiodic oscillation of SGR &  $L\gtrsim 50$ \\

Ref. \cite{BKA} & empirical approach+density functionals &  $64\pm5$ \\

Ref. \cite{FRDM}  & FRDM-2011a &  $70\pm15$   \\

Ref. \cite{GQR}  &giant quadrupole resonance energies &  $37 \pm 18$   \\

Ref. \cite{PDR}  &pygmy dipole resonance &  $64.8\pm15.7$  \\

Ref. \cite{Dong1}  &alpha-decay energies &  $61\pm22$  \\

Ref. \cite{AAA} & empirical approach, density functionals &  $59.0 \pm 13.0$ \\

Ref. \cite{dong13}  &beta-decay energies &  $50\pm15$  \\

Ref. \cite{charge}  &nuclear charge radius &  $54\pm19$   \\

Ref. \cite{Steiner}  & astrophysical observations of neutron star &  $43<L<52$   \\

Present  & nuclear mass differences &  $50.0\pm15.5$  \\
\end{tabular}
\end{ruledtabular}
\end{table}

The mass dependence of the symmetry energy coefficient
$a_{\text{sym}}(A)$ is given as \cite{DROP}
\begin{equation}
a_{\text{sym}}(A)=\frac{S_{0}}{1+\kappa A^{-1/3}},\label{F}
\end{equation}
where $\kappa$ is the ratio of the surface symmetry coefficient to
the volume symmetry coefficient. Centelles {\it et al}. proposed a
useful relation that the $a_{\text{sym}}(A)$ of finite nuclei is
approximately equal to $S(\rho _{A})$ of the nuclear matter at a
reference density $\rho_{A}$ \cite{MC}, which links the symmetry
energy of the nuclear matter and the one of finite nuclei, and thus
allows one to explore the density dependence of the symmetry energy
$S(\rho )$. The previous calculations showed that the reference
density $\rho _{A}\sim0.55\rho_{0}$ for $^{208}$Pb \cite{dong13},
where the model-dependence of the obtained $\rho_{A}$ in units of
$\rho_{0}$ is lowered greatly. The specific calculation process is
similar to that in Ref. \cite{dong13}. The formulism of DDM3Y shape
in Ref. \cite{OUR,TD0} is applied to describe the density dependence
of the symmetry energy $S(\rho)$
\begin{equation}
S(\rho )=13.0\left( \frac{\rho }{\rho _{0}}\right) ^{2/3}+C_{1}\left( \frac{%
\rho }{\rho _{0}}\right) +C_{2}\left( \frac{\rho }{\rho _{0}}\right)
^{5/3}. \label{DDDD}
\end{equation}
This formula is much more universal than the usually used
expressions $S(\rho )=S_{0}(\rho /\rho _{0})^{\gamma }$ and $S(\rho
)=12.5\left( \rho /\rho _{0}\right) ^{2/3}+C_{p}\left( \rho /\rho
_{0}\right) ^{\gamma }$ to describe the behavior of the symmetry
energy around the saturation density as pointed out in Ref.
\cite{OUR}, and it can provide both stiff and soft symmetry energy.

\begin{figure}[htbp]
\begin{center}
\includegraphics[width=0.45\textwidth]{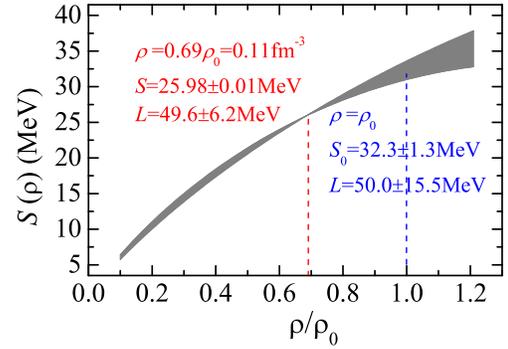}
\caption{(Color Online) Symmetry energy as a function of density.}
\end{center}
\end{figure}

Presently the symmetry energy $S_{0}$ at saturation density has been
determined relatively well, we solely determine the optimal value of
$\kappa$ (carrying error bars) taking the $S_{0}$ as an input. The
$S_{0}$ value has been constrained to rather narrow regions by some
authors, such as $S_{0}=32.5\pm0.5$ MeV from the mass systematics
\cite{FRDM}, $32.10\pm0.31$ MeV from the double differences of
¡°experimental¡± symmetry energies \cite{HJ}, $32.3\pm1.3$ MeV using
the PDR analysis combined with the correlation between $L$ and
$S_{0}$ \cite{PDR} and $31.5-33.5$ MeV from the calculations of
half-infinite matter \cite{DL}. The $S_{0}$ from the Carbone {\it et
al.} covering the other three is naturally believed to be the most
acceptable one \cite{dong13}. With their $S_{0}=32.3\pm1.3$ MeV as
input, the calculated value of $\kappa$ is $2.62^{+0.48}_{-0.46}$,
thus the symmetry energy coefficient of $^{208}$Pb is $22.4\pm0.3$
MeV. The parameters in Eq. (\ref{DDDD}) are $C_{1}=36.3\mp4.5$ MeV
and $C_{2}=-17.0\pm5.8$ MeV, and correspondingly the slope parameter
of nuclear symmetry energy is $L=50.0\pm15.5$ MeV. Incidently, if
the density dependent behavior $S(\rho )=S_{0}(\rho /\rho
_{0})^{\gamma}$ is applied, one obtains $L=59.3\pm 6.7$ MeV. The
reference density $\rho_{A}$ plays an important role to determine
the slope parameter $L$. If the widely used $\rho _{A}=0.1$
fm$^{-3}$ for $^{208}$Pb and $\rho _{0}=0.16$ fm$^{-3}$ are
employed, the obtained slope parameter is $L=74.0\pm17.5$ MeV. These
overestimate the $L$ value compared with the present calculations.
Table I shows the present estimated $L$ values compared with those
from other approaches. One can see clearly that the present finding
has a remarkable overlap with these results, in particular it is
very consistent with the later results, indicating that one may
relatively well understand the density-dependent behavior of
symmetry energy around the saturation density. The present approach
is much more straightforward than those applying the binding energy
directly. In this work, the effect of the Coulomb interaction on the
surface asymmetry and the effect of the surface diffuseness on the
Coulomb energy that described by $\Delta$ in Eq. (\ref{AA}) are
taken into account, which tend to be neglected in many previous
investigations. If we neglect these two effect, the estimated
$a_{\text{sym}}({A})$ of $^{208}\text{Pb}$ should be $22.9\pm 0.3$
MeV, and correspondingly the slope parameter is
$L=44.0^{+14.8}_{-15.6}$ MeV. Therefore, these effects cannot be
discarded optionally. Fig. 1 displays the density dependent behavior
of the symmetry energy versus density. It has been shown that the
neutron skin thickness of heavy nuclei is uniquely fixed by the
symmetry energy density slope $L(\rho)$ at a subsaturation cross
density $\rho \approx 0.11$fm$^{-3}$ rather than at saturation
density $\rho_{0}$ \cite{chen1}. And the giant monopole resonance of
heavy nuclei has been shown to be constrained by the EOS of nuclear
matter at $\rho \approx 0.11$fm$^{-3}$ rather than at saturation
density \cite{EMV}. Most interestingly, we find that at the density
of $\rho=0.69\rho_{0}=0.11$fm$^{-3}$, the symmetry energy is
$S(\rho=0.11\text{fm}^{-3})=25.98\pm0.01$ MeV, being agreement with
$26.2\pm1.0$ MeV that from the Skyrme forces in Ref. \cite{W2} and
$26.65\pm0.20$ MeV using data on neutron skin thickness of Sn
isotopes and binding energy differences for a number of heavy
isotope pairs \cite{chen1}. The small error bar results from the
fact that different curves almost intersect at this point as shown
in Fig. 1. Incidentally, if one use the reference density $\rho
_{A}=0.1$ fm$^{-3}$ for $^{208}$Pb, the obtained symmetry energy is
$S(\rho=0.11\text{fm}^{-3})=24.2\pm0.1$ MeV, which is lower than
that in the present calculation and in Ref. \cite{W2,chen1}. This
further suggests the importance of obtaining an accurate reference
density. In addition, the slope parameter is estimated to be
$L=49.6\pm6.2$ MeV, consistent with $L=46.0\pm4.5$ MeV in Ref.
\cite{chen1}. These results at $\rho \approx 0.11$fm$^{-3}$ are
perhaps useful to determine the neutron skin thickness with a higher
accuracy and to explore the cooling of canonical neutron stars.

In summary, the symmetry energy coefficients of heavy nuclei were
determined with the available experimental nuclear masses of heavy
nuclei. This approach prevents interferences from other energy terms
very effectively. The calculated symmetry energy coefficient of
$^{208}$Pb was furthermore employed to probe the density-dependent
symmetry energy of nuclear matter. With the symmetry energy
$S_{0}=32.3\pm1.3$ MeV at saturation density in Ref. \cite{PDR} as
an input, the estimated values of the slope parameter is
$L=50.0\pm15.5$ MeV, which agrees with very recent results, such as
these from the rms charge radius difference of the mirror nuclei
\cite{charge}, the $\beta^{-}$-decay energies of odd-A heavy nuclei
\cite{dong13} and astrophysical observations of neutron star
\cite{Steiner}. Moreover, we pay special attention to the symmetry
energy at the density of $\rho=0.69\rho_{0}=0.11$fm$^{-3}$ due to
its importance. The symmetry energy and the correspondingly slope
parameter are $S(\rho=0.11\text{fm}^{-3})=25.98\pm0.01$ MeV and
$L=49.6\pm6.2$ MeV respectively, which are consistent with few
published results. To reduce the uncertainty of the $L$ value, one
need to reduce the uncertainty of the $S_{0}$ value as far as
possible.

This work was supported by the 973 Program of China under Grants No.
2013CB834405, the National Natural Science Foundation of China under
Grants No. 11175219, 10975190, and 11275271; the Knowledge
Innovation Project (KJCX2-EW-N01) of Chinese Academy of Sciences.

\end{document}